\documentclass{ws-ijmpcs}
\usepackage{slashed}
\usepackage{amssymb}

\usepackage{ulem}
\usepackage{color}

\newcommand{\Eqref}[1]{Eq.~\eqref{#1}}

\begin{document}

\markboth{Felix Karbstein, Lars Roessler, Babette D\"obrich, Holger Gies}
{Probing the quantum vacuum: The photon polarization tensor}

%
\catchline{}{}{}{}{}
%

\title{OPTICAL PROBES OF THE QUANTUM VACUUM:\\ THE PHOTON POLARIZATION TENSOR IN EXTERNAL FIELDS}

\author{FELIX KARBSTEIN, LARS ROESSLER, BABETTE D\"OBRICH, HOLGER GIES}

\address{
Helmholtz Institut Jena, Fr\"obelstieg 3, D-07743 Jena, Germany, and
Theoretisch-Physikalisches Institut, 
Friedrich-Schiller-Universit\"at Jena, Max-Wien-Platz 1,
D-07743 Jena, Germany\\
\{felix.karbstein,lars.roessler,babette.doebrich,holger.gies\}@uni-jena.de}

\maketitle

\begin{history}
\end{history}

\begin{abstract}
The photon polarization tensor is the central building block of an effective
theory description of photon propagation in the quantum vacuum.  It accounts
for the vacuum fluctuations of the underlying theory, and in the presence of
external electromagnetic fields, gives rise to such striking phenomena as
vacuum birefringence and dichroism. Standard approximations of the polarization tensor are
often restricted to on-the-light-cone dynamics in homogeneous electromagnetic fields, and are limited to certain momentum regimes only.
We devise two different strategies to go beyond these limitations: First, we
aim at obtaining novel analytical insights into the photon polarization tensor
for homogeneous fields, while retaining its full momentum dependence.  Second,
we employ wordline numerical methods to surpass the constant-field limit.

\keywords{Quantum vacuum; external fields; photon polarization tensor.}
\end{abstract}

\ccode{PACS numbers: 11.25.Hf, 123.1K}

\section{Introduction}	

The photon polarization tensor is a central quantity in quantum
electrodynamics (QED). {It contains essential information about the
  renormalization properties of QED and encodes quantum corrections to
  Coulomb's force law.}  Accounting for the vacuum fluctuations which modify the
propagation of photons, the polarization tensor probes the particle content of
the underlying theory{, such as QED or even beyond}. The effective
theory for soft electromagnetic fields in the quantum vacuum is provided
  by the famous Heisenberg-Euler Lagrangian\cite{physics/0605038}. For photon
propagation at arbitrary frequencies, the generalization reads (reviewed, e.g., in
Ref.~\refcite{Dittrich:2000zu}),
\begin{equation}
\mathcal{L}[{\mathcal{A}}]= -\frac{1}{4} {\mathcal{F}}_{\mu\nu}(x) 
{\mathcal{F}}^{\mu\nu}(x) 
- \frac{1}{2}\int_{x'}  a_\mu(x) \Pi^{\mu\nu}(x,x') a_\nu(x')\,,\label{eq:calL}
\end{equation}
with $\Pi^{\mu\nu}(x,x')$ denoting the photon polarization tensor.  Here,
$\mathcal{F}_{\mu \nu}$ is the field strength tensor of a classical,
  macroscopic field $\mathcal{A}_\mu=A_\mu+a_\mu$ which we decompose into a
  propagating photon field $a_\mu$ with an amplitude that is considered to be
  weak compared to the electron mass scale and an external potentially
    strong electromagnetic field $A_\mu$. The polarization tensor
$\Pi^{\mu\nu}(x,x')$, being the second-order correlation function of the
  photon field, acquires a dependence on this external field $A_\mu$ and thus
  modifies the propagation of photons in the vacuum.  Hence, the photon polarization tensor acts as the source for exciting effects such as vacuum
birefringence and
dichroism\cite{Toll:1952}\cdash\cite{Adler:1971wn}. As 
it is sensitive to all charged fluctuations in the vacuum,
  it can also be used as a probe for exotic particles beyond the standard
  model, such as minicharged degrees of freedom coupling to
  electromagnetism\cite{Holdom:1985ag,Gies:2006ca}.

Whereas for homogeneous electromagnetic fields, the photon polarization tensor at one-loop level is known exactly in terms of
a double parameter integral in momentum space since a long time\cite{BatShab}
(cf. also Ref.~\refcite{Dittrich:2000zu}, and references therein), its
explicit evaluation still poses an intricate issue.  Basically all insights
available so far have been derived from this expression by means of various
approximation techniques.  Apart from the standard perturbative weak-field
expansion, these approximations allow for insights in particular strong-field regimes also.  However, the latter approximations put special attention to
on-the-light-cone dynamics, and are in general limited to physical problems
that can be treated directly in momentum space.  This is due to the fact that
their derivation involves constraints to a certain momentum regime, whereas
the transition to position space by a Fourier transformation requires
knowledge about the full momentum range.

Hence, recent advances in the field of laser physics, as well as growing
interest in the experimental search for beyond-the-standard-model particles,
like minicharges, strongly necessitate new insights into the photon
polarization tensor in the presence of external electromagnetic fields. On the
one hand, this is essential for novel studies in the framework of
QED which come into reach owing to the fast advances in the field of high-intensity laser physics\cite{DiPiazza:2011tq}. On the other hand, for minicharged particles,
neither their mass, nor their charge is restricted {\it a priori}, which requires
knowledge about the full parameter regime, particularly also in the strong-field
limit. 
Particular\cite{Gies:2009wx,DGNK}
light-shining-through-walls type experiments\cite{Redondo:2010dp} also require knowledge about 
the full momentum dependence.  
Similar considerations apply to inhomogeneous fields, as soon as the
  scale of temporal or spatial variations of the external field becomes
  comparable to the scale set by the Compton wavelength of
the virtual particles. In the case of QED,
  the superposition of a strong optical laser field with higher harmonics in
  the X-ray or gamma-ray regime can lead to strong violations of the
  homogeneous-field assumption. Beyond QED, minicharges can correspond to very
  large or even macroscopic Compton wavelengths, so that inhomogeneous-field
  configurations represent the standard rather than the exceptional case.

Here we present some attempts to tackle these problems.  First, we provide analytical access to the photon polarization tensor in the
presence of a strong, homogeneous magnetic field, while keeping its full
momentum dependence.  Second, we employ wordline numerical methods to surpass
the constant-field limit, and in particular study birefringence in
a spatially inhomogeneous magnetic field.

\section{The photon polarization tensor}

We focus on the photon polarization tensor at one-loop level.  Whereas it is
known exactly for arbitrary homogeneous, externally set electromagnetic field
configurations in terms of a double parameter
integral\cite{Dittrich:2000zu,BatShab,Urrutia:1977xb,Schubert:2000yt}, we here limit
ourselves to the special case of a purely magnetic
field\cite{Tsai:1974fa,Tsai:1975iz}. Hence, the only two externally set vectors in the
problem are the magnetic field $\vec{B}$ and the wave vector of the probe
photons. They govern the entire direction dependence of the polarization
tensor. Of course, in inhomogeneous fields, the tensor structure can become much more involved.

\subsection{The basic setting}

In the constant magnetic field situation, four-vectors $k^{\mu}$ are then naturally
decomposed into components parallel and perpendicular to the magnetic field
vector $\vec B$.  Without loss of generality, $\vec B$ is assumed to point in
${\vec e}_1$ direction, and the following decomposition\cite{Urrutia:1977xb} is adopted,
\begin{align}
 k^{\mu}=k_{\parallel}^{\mu}+k_{\perp}^{\mu}\,,\quad\quad k_{\parallel}^{\mu}=(k^0,k^1,0,0)\,,\quad\quad k_{\perp}^{\mu}=(0,0,k^2,k^3)\,. \label{eq:k_dec}
\end{align}
In the same manner tensors can be decomposed, e.g.,
$g^{\mu\nu}=g_{\parallel}^{\mu\nu}+g_{\perp}^{\mu\nu}$.  It is then convenient
to introduce projection operators,
\begin{align}
 &P^{\mu\nu}_{1}=g_{\parallel}^{\mu\nu}-\frac{k_{\parallel}^{\mu}k_{\parallel}^{\nu}}{k_{\parallel}^2}\ , \quad\quad
 P^{\mu\nu}_{2}=g_{\perp}^{\mu\nu}-\frac{k_{\perp}^{\mu}k_{\perp}^{\nu}}{k_{\perp}^2}\ , \nonumber\\
 &P^{\mu\nu}_{3}=g^{\mu\nu}-\frac{k^{\mu}k^{\nu}}{k^2}-P^{\mu\nu}_{1}-P^{\mu\nu}_{2}\ .
\label{eq:Projs}
\end{align}
We use a metric with signature $(-,+,+,+)$, i.e., $k^2=\vec{k}^2-(k^0)^2$.
For a given photon four-momentum $k^{\mu}$, the projectors $P^{\mu\nu}_{p}$
($p=1,2,3$) project onto the three independent photon polarization modes in
the presence of an external field.  As the vacuum speed of light in external
fields deviates from its zero-field value, and the vacuum exhibits medium-like
properties, the occurrence of three (instead of two in the absence of an
external field) independent polarization modes is not surprising.

As long as $\vec{k}\nparallel\vec{B}$, the projectors $P^{\mu\nu}_{1}$ and
$P^{\mu\nu}_{2}$ have an intuitive interpretation. They project onto photon
modes polarized in parallel and perpendicular to the plane spanned by the two
vectors $\vec{k}$ and $\vec{B}$. For $\vec{k}\nparallel\vec{B}$ these are the
polarization modes that can be continuously related to those in the
limit of vanishing external field.  For the special alignment of $\vec{k}\parallel\vec{B}$ only
one externally set direction is left, and we encounter rotational invariance
arround the magnetic field axis. Here, the modes $2$ and $3$ can be
continuously related to the two zero-field polarization modes.

\subsection{The photon polarization tensor for homogeneous fields}

We consider the standard QED polarization tensor induced by vacuum fluctuations of charged Dirac fermions, and use its proper-time\cite{Schwinger:1951nm} representation in momentum space. As long as the magnetic field is homogeneous,
translational invariance implies $\Pi^{\mu\nu}(x,x')=\Pi^{\mu\nu}(x-x')$ and
the polarization tensor in momentum space depends on the single momentum
$k^{\mu}$ only.  It is of the general form\cite{Dittrich:2000zu}
\begin{align}
 \Pi^{\mu\nu}(k)=\Pi_{1}(k)P^{\mu\nu}_{1}+\Pi_{2}(k)P^{\mu\nu}_{2}+\Pi_{3}(k)P^{\mu\nu}_{3}\,, \label{eq:PI}
\end{align}
where the scalar functions $\Pi_p(k)$ ($p=1,2,3$) are the components of the polarization tensor in the respective subspaces.
Their explicit expressions read
\begin{align}
\left\{
 \begin{array}{c}
 \Pi_{1}\\
 \Pi_{2}\\
 \Pi_{3}
 \end{array}
\right\}
=\frac{\alpha}{2\pi}\int\limits_0^{\infty}\frac{{\rm d} s}{s}\int\limits_{0}^{1}{\rm d}\nu\left[{\rm e}^{-{\rm i}\Phi_0s}\frac{z}{\sin z}
\left(
\left\{
 \begin{array}{c}
 \tilde{N}_1 \\
 N_0\\
 N_0
 \end{array}
\right\}k_{\parallel}^2
+
\left\{
 \begin{array}{c}
 N_0 \\
 \tilde{N}_2 \\
 N_0
 \end{array}
\right\}k_{\perp}^2
\right) + {\rm c.t.}
\right], \label{eq:PI_comp}
\end{align}
with contact term
\begin{equation}
 {\rm c.t.}=-(1-\nu^2)\,{\rm e}^{-{\rm i}\left(m^2-{\rm i}\epsilon\right)s}\,k^2\,.
\end{equation}
The dependence on $B=|\vec{B}|$ is encoded in the variable $z=eBs$, $\epsilon>0$ denotes an infinitesimal parameter, $\alpha=e^2/(4\pi)$ is the the fine-structure constant, and
\begin{align}
 \Phi_0&=m^2-{\rm i}\epsilon+\frac{1-\nu^2}{4}{k}_{\parallel}^2+\frac{\cos{\nu z}-\cos{z}}{2z\sin{z}}{k}_{\perp}^2\,, \label{eq:Phi0} \\
 N_0&=\cos\nu z-\nu\sin \nu z \cot z\,, \nonumber\\
 \tilde{N}_1&=(1-\nu^2)\cos z\,, \nonumber\\
 \tilde{N}_2&=2\frac{\cos \nu z -\cos z}{\sin^2z}\,.
\end{align}
The parameter $s$ denotes the propertime, and $\nu$ governs the momentum distribution within the loop.
In particular due to the $s$-dependence of the phase factor, Eq.~(\ref{eq:Phi0}), via trigonometric functions, the propertime integral in general cannot be performed analytically, and is also hard to tackle numerically.

\subsubsection{Approximations to the polarization tensor}

Basically all explicit insights into the photon polarization tensor in the presence of a constant magnetic field can be traced back to three well-established classes of approximations:
\begin{itemize}
 \item a perturbative expansion in the number of external field insertions in the particle-antiparticle loop, which can be associated with the limit $\frac{eB}{m^2}\ll1$,
 \item a quasi-classical approximation\cite{Baier:2009it} developed in the seminal works of Tsai and Erber\cite{Tsai:1974fa,Tsai:1975iz}, derived ``on-the-light-cone'', 
i.e., for $k^2=0$, and restricted to $\frac{k^2_{\perp}}{eB}\gg1$ only, and
 \item the restriction to the lowest Landau level, or equivalently a ``large-$z$'' expansion\cite{Shabad:1975ik}, valid in the limit where $\frac{eB}{m^2}\gg1$, 
and so-far commonly utilized below pair-creation threshold, $\omega^2<4m^2$. 
\end{itemize}

\subsubsection{The special alignment $\vec k\parallel\vec B$}

In order to go beyond these approximation schemes, we consider
 the situation $\vec k\parallel\vec B$, i.e., $k_{\perp}^{\mu}=0$ and
 $k^{\mu}= k_{\parallel}^{\mu}$.  
In this limit the $z$-dependence in Eq.~(\ref{eq:Phi0}) drops out, and the propertime integration simplifies significantly.

\begin{figure}[ht]
\centerline{\includegraphics[width=2cm]{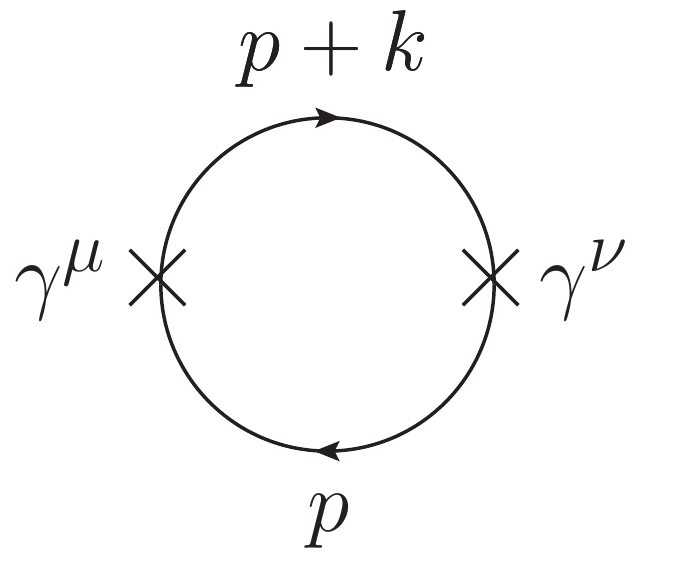}}
\caption{One-loop diagram corresponding to the photon polarization tensor in the absence of an external field. The solid lines represent Dirac fermions.\label{fig:karbstein1}}
\end{figure}

It is illustrative to focus on mode $1$, even though this is the mode that
cannot be continuously related to a zero-field polarization mode. For $k_{\perp}^{\mu}=0$ the component $\Pi_1(k)$ can easily be evaluated following
an alternative approach also, and hence provides a useful means to unambiguously
fix the propertime integration contour in Eq.~(\ref{eq:PI_comp}).  The reason
for this is twofold: The projector $P_1^{\mu\nu}$ is completely independent of
$k^{\mu}_{\perp}$, and in the presence of a magnetic field we encounter Landau
level quantization for momentum components perpendicular to the magnetic field
vector $\vec B$.  In the absence of external fields the photon polarization
tensor is easily determined by evaluating the loop diagram depicted in
Fig.~\ref{fig:karbstein1}.
In $D=d+1$ space-time dimensions we obtain for the unrenormalized polarization tensor
\begin{eqnarray}
 \Pi^{\mu\nu}(k)&=&{\rm i}({\rm i}e)^2\,{\rm tr}\left\{\int\frac{{\rm d}^Dp}{(2\pi)^D}\,\gamma^{\mu}\frac{\rm i}{\slashed{p}-m-{\rm i}\epsilon}\gamma^{\nu}\frac{\rm i}{\slashed{p}+\slashed{k}-m-{\rm i}\epsilon}\right\} \label{eq:PI_Ddim}\\
&=&\left(k^2g^{\mu\nu}-k^{\mu}k^{\nu}\right)\frac{\alpha D}{2}\,\frac{\Gamma\left(\frac{4-D}{2}\right)}{(4\pi)^{\frac{D-2}{2}}}\int_0^1 {\rm d}\nu\,(1-\nu^2)\,\left[\frac{1}{m^2-{\rm i}\epsilon+\frac{1-\nu^2}{4}k^2}\right]^{\frac{4-D}{2}}. \nonumber
\end{eqnarray}
The trace is over Dirac indices.
For the $1$-mode, and in the limit $k_{\perp}^{\mu}=0$, this implies
\begin{equation}
 \Pi_1(k)=k_{\parallel}^2\frac{\alpha D}{2}\,\frac{\Gamma\left(\frac{4-D}{2}\right)}{(4\pi)^{\frac{D-2}{2}}}\int_0^1 {\rm d}\nu\,(1-\nu^2)\,\left[\frac{1}{m^2-{\rm i}\epsilon+\frac{1-\nu^2}{4}k_{\parallel}^2}\right]^{\frac{4-D}{2}}. \label{eq:PI_1Ddim}
\end{equation}
Turning to $D=3+1$ dimensions and imposing gauge-invariant renormalization
conditions, the polarization tensor vanishes for photon wave vectors on
the light cone. This results in the familiar expression for vanishing external fields
\begin{equation}
 \left.\Pi_1(k)\right|_{B=0}=(k_{\parallel}^2)^2\frac{\alpha}{4\pi}\int_0^1{\rm d}\nu\left(\frac{\nu^2}{3}-1\right)\frac{\nu^2}{m^2-{\rm i}\epsilon+\frac{1-\nu^2}{4}k_{\parallel}^2}\ .\label{eq:PI_vac}
\end{equation}

In order to evaluate $\Pi_1(k)$ in the presence of a magnetic field $\vec B\parallel\vec k$ in $D=3+1$ dimensions, 
we rewrite the integral over the loop momentum in Eq.~(\ref{eq:PI_Ddim}) as follows,
\begin{align}
 \int\frac{{\rm d}^4p}{(2\pi)^4}=\int\frac{{\rm d}^2p_{\parallel}}{(2\pi)^2}\int\frac{{\rm d}p^2_{\perp}}{4\pi}\,. \label{eq:int_measure1}
\end{align}
Landau level quantization, implies
\begin{equation}
 p_{\perp}^2=2eBn,\quad {\rm with}\ n\in\mathbb{N}_0\,,
\end{equation}
and results in
\begin{align}
 \int\frac{{\rm d}^4p}{(2\pi)^4}\quad\to\quad\frac{eB}{2\pi}\sum_{n=0}^{\infty}c_n\int\frac{{\rm d}^2p_{\parallel}}{(2\pi)^2}\,. \label{eq:int_measure2}
\end{align}
The multiplicity factor $c_n$ accounts for spin degrees of freedom. It is $1$ for $n=0$, but $2$ for $n\in\mathbb{N}$.
We then perform the residual momentum integral in $D=1+1$ dimensions, cf. Eq.~(\ref{eq:int_measure2}), substitute
\begin{equation}
 m^2\quad\to\quad m_n^2=m^2+2eBn,
\end{equation}
and obtain
\begin{eqnarray}
 \Pi_{1}(k)&=&k_{\parallel}^2\frac{\alpha eB}{2\pi}\int_0^1{\rm d}\nu\,
(1-\nu^2)\,\sum_{n=0}^{\infty}\frac{c_n}{m_n^2-{\rm i}\epsilon+\frac{1-\nu^2}{4}k_{\parallel}^2}\,. \label{eq:PI_Landau}
\end{eqnarray}
After renormalization such that Eq.~(\ref{eq:PI_vac}) is retained in the zero-field limit, Eq.~(\ref{eq:PI_Landau}) can be cast in the following concise form\cite{BF},
\begin{equation}
 \Pi_{1}(k)=k_{\parallel}^2\frac{\alpha}{2\pi}\int_0^1 {\rm d}\nu\,
(1-\nu^2)\left[\ln\left(\frac{m^2-{\rm i}\epsilon}{2eB}\right)-\Psi\left(\frac{\left.\Phi_0\right|_{\parallel}}{2eB}\right)-\frac{eB}{\left.\Phi_0\right|_{\parallel}}\right], \label{eq:PI_parallelexact}
\end{equation}
with (cf. Eq.~(\ref{eq:Phi0}))
\begin{equation}
 \left.\Phi_0\right|_{\parallel}=m^2-{\rm i}\epsilon+\frac{1-\nu^2}{4}{k}_{\parallel}^2\,.
\end{equation}
Here we made use of the exact series representation of the Digamma function\cite{Gradshteyn},
\begin{equation}
 \Psi(\xi)=-\gamma-\frac{1}{\xi}+\sum_{n=1}^{\infty}\frac{\xi}{n(\xi+n)}\,,
\end{equation}
where $\gamma$ denotes the Euler-Mascheroni constant.

Let us emphasize that Eq.~(\ref{eq:PI_parallelexact}) is valid in the full momentum regime, i.e., in particular also beyond the pair creation threshold. Besides the special alignment $\vec k\parallel\vec B$, the above derivation does not involve any further restrictions.
In particular, the leading contribution of Eq.~(\ref{eq:PI_parallelexact}) in the limit $B\to\infty$ is linear in $B$ and reads
\begin{equation}
 \Pi_{1}(k)\ \xrightarrow{B\to\infty}\ k_{\parallel}^2\ \frac{\alpha eB}{2\pi}\int_0^1{\rm d}\nu\ \frac{1-\nu^2}{m^2-{\rm i}\epsilon+\frac{1-\nu^2}{4}k_{\parallel}^2}\ . \label{eq:PI_LOlargeB}
\end{equation}

Of course Eq.~(\ref{eq:PI_parallelexact}), derived by summing over all Landau levels in the magnetic field, should also be reproduced when performing the respective propertime integral in Eq.~(\ref{eq:PI_comp}) directly.
It turns out that this is indeed so, if the propertime integration contour in Eq.~(\ref{eq:PI_comp}) lies slightly below the positive real axis (cf. also Ref.~\refcite{Baier:2009it}), i.e.,
\begin{equation}
 \int_0^{\infty}{\rm d}s \quad\to\quad\lim_{\eta\to0}\,\int_{0-{\rm i}\eta}^{\infty-{\rm i}\eta}{\rm d}s\,. \label{eq:contour}
\end{equation}
Only then the propertime integral in Eq.~(\ref{eq:PI_comp}) is well
defined. Given that $\vec k \parallel \vec B$, this prescription of the
integration contour allows us to perform the propertime integral
explicitly\cite{BF} for all the components $\Pi_p(k)$ ($p=1,2,3$) of the
photon polarization tensor; see also Refs.~\refcite{Cover:1974ij} and \refcite{Tsai:1975tw}. Note that
$\Pi_2(k)=\Pi_3(k)$ in this limit.  Instead of the double parameter integral
representation, we are left with a single parameter integral, spanning a
finite integration interval only.  Let us emphasize again, that the resulting
expressions are indeed valid in the full momentum regime. Hence, they are
amenable to a Fourier transformation, and can in principle also be employed in
position space.

\subsubsection{The situation $\vec k\nparallel\vec B$}

For $\vec k\nparallel\vec B$ the propertime integration in general cannot be
performed explicitly. However, also here we obtain insights into the strong-field
regime. By employing an analytical continuation $eB\to-{\rm i}eB$, which is
permissible as the integration contour lies below the real axis, see
Eq.~(\ref{eq:contour}) (cf. Ref.~\refcite{BF} for a detailed discussion), we
can easily extract the leading contribution in the strong-field limit 
$B\to\infty$.  We obtain
\begin{equation}
 \Pi^{\mu\nu}(k)\ \xrightarrow{B\to\infty} P^{\mu\nu}_{1}\ k_{\parallel}^2\ \frac{\alpha eB}{2\pi}\ 
{\rm e}^{-\frac{k_{\perp}^2}{2eB}}\int_{0}^{1}{\rm d}\nu\  \frac{1-\nu^2}{m^2-{\rm i}\epsilon+\frac{1-\nu^2}{4}k_{\parallel}^2}\ . \label{eq:PI_largez}
\end{equation}
Note, that the leading contribution in the strong-field limit arises from $\Pi_1(k)$, whereas $\Pi_2(k)$ and $\Pi_3(k)$ are suppressed and start contributing at subleading order only\cite{BF}. 
 Moreover, a factorization with respect to the momentum dependence,  $k_{\parallel}^{\mu}$ and $k_{\perp}^{\mu}$, is encountered here. In the limit $k_{\perp}\to0$, 
Eq.~(\ref{eq:PI_LOlargeB}) is retained.
Eq.~(\ref{eq:PI_largez}) provides information about the truly non-perturbative regime in the situation where $\vec k\nparallel\vec B$. It therefore is of 
particular interest in attempts to restrict the available parameter regime for beyond-the-standard-model particles, such as minicharges\cite{DGNK}.

\subsection{The photon polarization tensor for inhomogeneous fields}

When going beyond the constant-field approximation, the standard tools
based on a Feynman diagrammatic language become inefficient as they
technically require the diagonalization of Laplace-type operators in general
inhomogeneous backgrounds. Each background configuration thus represents a new
computational challenge. An elegant way to circumvent this problem is provided
by the worldline approach\cite{Schubert:2001he}, where traces over Laplacian
operators are rewritten in terms of Feynman path integrals in position
space. In fact, worldline expressions for correlation functions to arbitrarily
high order can be written down in closed form within perturbation theory. As
these formulas ultimately require to carry out a path integral, powerful
Monte-Carlo algorithms\cite{Gies:2001zp}\cdash\cite{Gies:2003cv}
can be used to reliably extract quantitative information in rather general
background fields. Whereas the general formalism can straightforwardly be
extended to spinor QED\cite{Langfeld:2002vy}, we here confine ourselves to
scalar QED (with spinless electrons). 

\subsubsection{Worldline representation}

A starting point of the formalism is given by the one-loop effective action in
$D$ Euclidean spacetime dimensions\cite{Schubert:2001he},
\begin{align}
	\Gamma[\mathcal{A}] =
        \int\limits_{0}^{\infty}\frac{dT}{T}\frac{e^{-m^2T}}{(4\pi
          T)^{D/2}}\int\limits_{x(0) =
          x(T)}\mathcal{D}x~e^{-\int_{0}^{T}d\tau~(\frac{\dot{x}^2}{4}+{\rm i}e\dot{x}\mathcal{A})} ,
\end{align}
being the generating functional of all 1-loop 1PI correlation functions of
the photon field. The polarization tensor is contained in the expansion of
$\Gamma[\mathcal{A}]$ to second order ($j=2$) in a propagating photon field, i.e.,
$\mathcal{A}_{j}^{\mu}(x) = {A}^\mu+ \sum\limits_{j =
  1}^{\infty}\epsilon_{j}^{\mu}e^{ik_{j}x}$, where ${A}^{\mu}$ denotes the
external background field. This second-order expansion exactly matches with
\Eqref{eq:calL} with $\Gamma= \int d^D x \mathcal{L}$. 

A new feature of inhomogeneous fields is that translational invariance is
generically broken, resulting in the fact that the polarization tensor depends
on incoming and outgoing momenta independently. For simplicity, we here
confine ourselves to the case where the field is translationally invariant
along the direction of photon propagation, such that the photon momentum
remains conserved. The formalism for the general case is described in
Ref.~\refcite{Gies:2011he}. In the special case, the worldline expression for
the unrenormalized polarization tensor reads ($P_\text{T}^{\mu\nu}=g^{\mu\nu}
-k^\mu k^\nu/k^2$) 
\begin{eqnarray}
\Pi^{\mu\nu}(k) 
& =& \frac{(-{\rm i}e)^2}{(4\pi)^{D/2}} 
\int\limits_{0}^{\infty}\frac{dT}{{T^{D/2}}}{e^{-m^2T}}
P_{\text{T}}^{\mu\kappa}(k)P_{\text{T}}^{\nu\lambda}(k)\label{eq:Pivac} \\
&&\times \left\langle\int\limits_{0}^{1}d\tau_{1}\int\limits_{0}^{1}d\tau_{2}~
            \dot{y}_{1,\kappa}  e^{{\rm i}\sqrt{T}ky_{1}}
          \dot{y}_{2,\lambda}
        e^{-{\rm i}\sqrt{T}ky_{2}} e^{-{\rm i}e\sqrt{T}\oint dy\cdot{A}(x_{\text{CM}}+ \sqrt{T} y))}\right\rangle, \nonumber
\end{eqnarray}
where the expectation value is defined via a propertime-rescaled integral over
worldlines $y_\mu(\tau)$ centered around a common center of mass
$x_{\text{CM}}$, see Ref.~\refcite{Gies:2011he} for details.

In the absence of an external field $A^\mu=0$, the standard vacuum polarization
tensor for scalar QED analogous to \Eqref{eq:PI_Ddim} is recovered
analytically as well as numerically\cite{Gies:2011he}. Also the analytical
scalar QED result for the polarization tensor in homogeneous magnetic
fields\cite{Schubert:2000yt} can very well be reproduced by the numerical
algorithm for weak as well as strong fields.

\subsubsection{Light propagation in inhomogeneous fields}

Whereas the numerical worldline formalism is primarily constructed in
Euclidean space as dictated by the Monte Carlo importance
sampling, it is nevertheless possible to insert Minkowski-valued 4-momentum
vectors $k_\mu=({\rm i} \omega, \vec{k})$. The latter is in fact necessary in
order to extract light propagation properties in Minkowski space. 

In the following, we concentrate on the birefringence properties of the
magnetized quantum vacuum and consider the case of a propagation direction
orthogonal to the magnetic field, $\vec{k}\perp\vec{B}$. In this case,
the two different polarization modes corresponding to the projectors
$P_1^{\mu\nu},P_2^{\mu\nu}$ in \Eqref{eq:Projs} propagate at different phase velocities, $v=
\omega/|\vec{k}|$, 
\begin{align}\label{eq:phasevelocity}
v_{\| / \perp}^2 = 1 - \frac{\Pi_{1,2}}{|\vec{k}|^2} = (1- \Delta v_{\|/\perp})^2.
\end{align}
The weak-field limit of these velocity shifts is given by\cite{Ahlers:2006iz} 
\begin{align}
\Delta v_{\| / \perp} =a_{\|/\perp}
\frac{\alpha}{4\pi}\frac{(eB)^2}{m^4}, \quad a_{\|/\perp}= \left\{
\begin{matrix}
\frac{1}{90} \\\\
\frac{7}{90}
\end{matrix}\right\}.
\label{eq:vshift}
\end{align}
This limit can be straightforwardly reproduced by worldline Monte Carlo
with a precision level of a few percent with moderate numerical
cost\cite{Gies:2011he}. 

\begin{figure}[t] 
\centerline{\includegraphics[width = 0.6\linewidth]{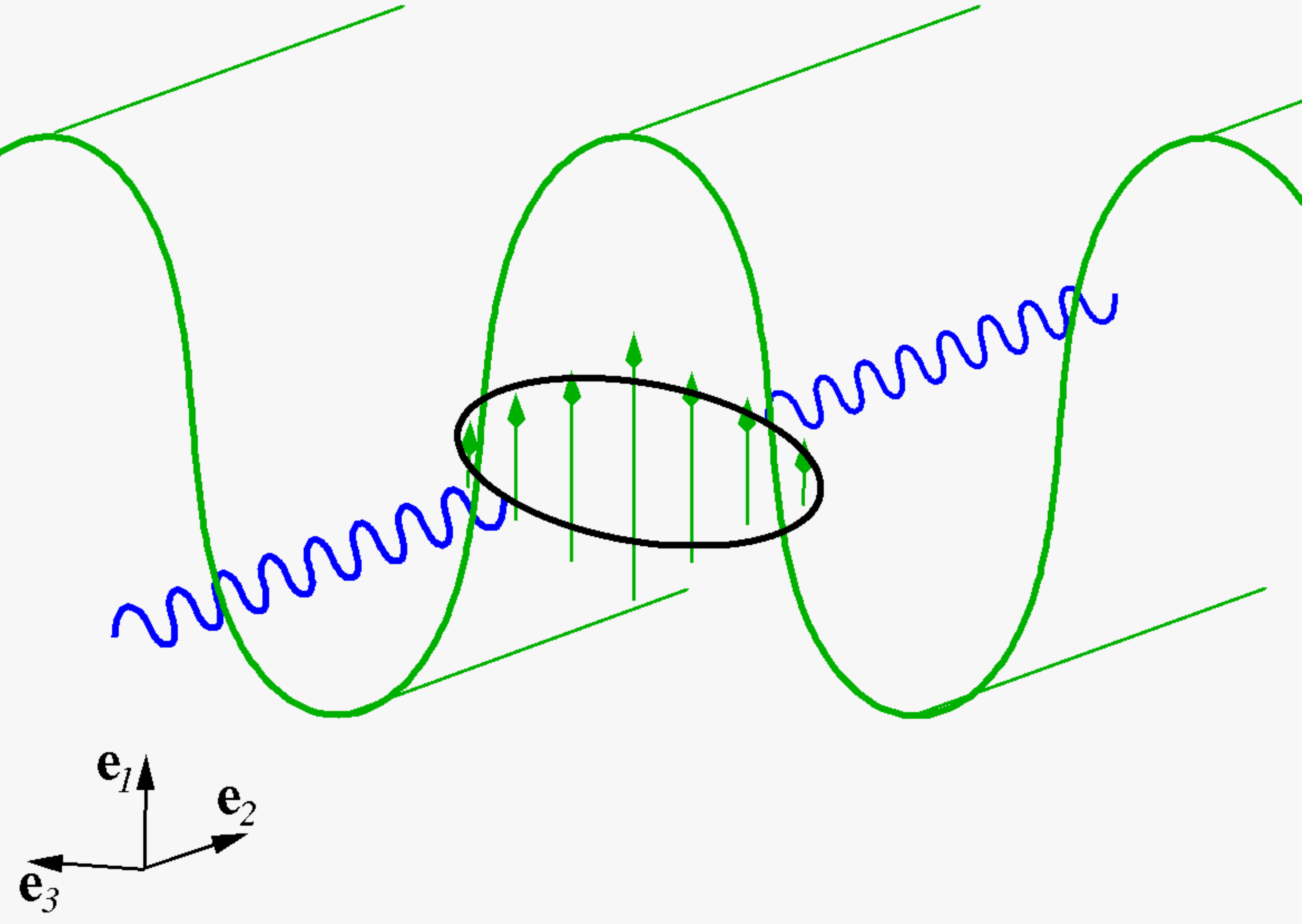}}
\caption{Sketch of the photon propagation in an inhomogeneous field
configuration \Eqref{eq:Bvary}.}
\label{fig:sketch}
\end{figure}

Let us here concentrate on new polarization effects in inhomogeneous fields,
characteristic for nonlocal features of fluctuation phenomena. For this, we
consider the superposition of a constant magnetic field $\overline{B}$
and a sinusoidal magnetic oscillation varying in $\vec{e}_3$ direction
with amplitude $B_1$ and wavelength $\lambda_B$,
\begin{equation}\label{eq:Bvary}
\vec{B}(x_3) = \left[\overline{B} +
  B_{1}\cos\left(\frac{2\pi}{\lambda_{B}}x_{3}\right)\right]\vec{e}_{1}.
\end{equation}
A sketch of the geometry is shown in Fig. \ref{fig:sketch}. 
This field
configuration is inspired by the superposition of a strong optical
standing-wave laser pulse and higher harmonics in the X- or gamma-ray regime. 

\begin{figure}[t]
\centerline{\includegraphics[scale = 0.18]{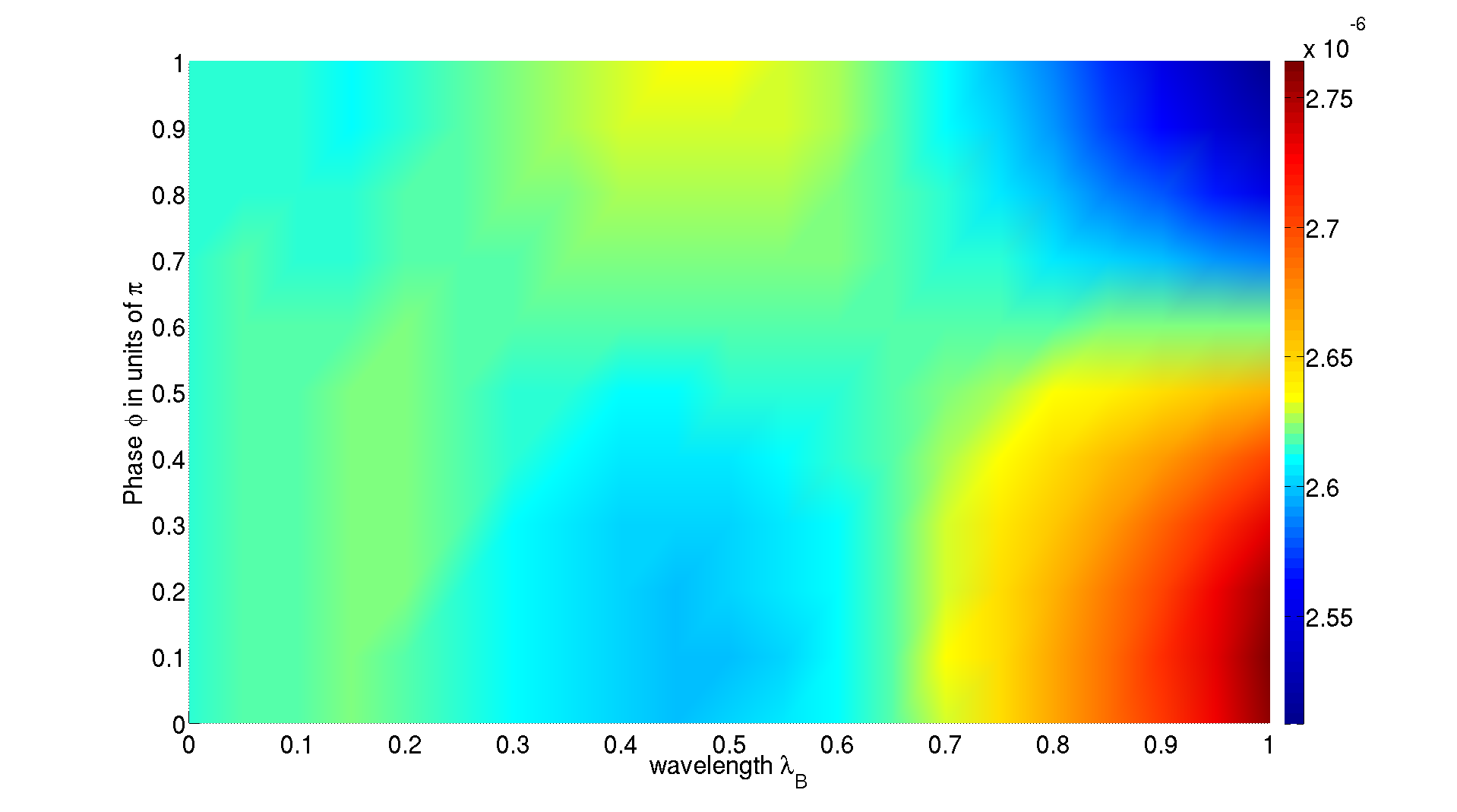} }
\caption{Contour plot of the phase velocity shift $\Delta v_\|$ for photon
  propagation in an inhomogeneous magnetic field \eqref{eq:Bvary} along the
  $\vec{e}_2$ direction as a function of the field variation wavelength
  $\lambda_B$ and the position phase $\phi=2\pi x_3/\lambda_B$. }
\label{fig:map}
\end{figure}

The velocity shifts now depend on the $x_3\equiv x_{3, \text{CM}}$ coordinate,
$\Delta v(x_3)$, given in terms of the local $x_3$ dependent eigenvalues of
the polarization tensor
\begin{equation}\label{equation_phasevelocity_in}
 v_{\| / \perp}^2(x_{3}) = 1 - \frac{\Pi_{1,2}(x_{3})}{|\vec{k}|^2}.
\end{equation}
In our numerical computations, we use $e\overline{B}=0.2 m^2$ and $B_{1} = 0.5
\overline{B}$. We study the dependence of the velocity shift $\Delta v_\|$ as
a function of the magnetic oscillation wavelength $\lambda_B$ and express the
position $x_3$ inside the magnetic oscillation in terms of a phase $\phi=2\pi
x_{3}/\lambda_B$; $\phi=0,2\pi,\dots$ corresponding to photon propagation along
the field maxima and $\phi=\pi, 3\pi,\dots$ to minima. In the limit of large
$\lambda_B\gg 1/m$, the field becomes slowly varying with respect to the
Compton wavelength. Here, the local velocity shifts approach the homogeneous
field limits \eqref{eq:vshift} upon insertion of the local magnetic field
$B(x_3)$.  Near $\lambda_B m\simeq 1$, the field oscillates on the scale of
the Compton wavelength, and larger deviations from the
``locally-constant-field'' approximation become visible, see
Fig.~\ref{fig:map}.
In the worldline picture, the propagating photon undergoing a virtual (scalar)
electron-positron loop with spatial extent $\sim 1/m$ recognizes a magnetic 
field averaged over the size of a Compton wavelength. This primarily leads to
a wash out of the velocity-shift contour with respect to the field
inhomogeneities. In the limit of very rapid variations, $\lambda_B \ll 1/m$,
the photon thus undergoes a velocity shift induced by the averaged field value
$\overline{B}$, such that $\Delta v \sim \overline{B}^2$.

An interesting observation in the region $\lambda_B m\lesssim 1$ is that the
transition from the locally-constant-field limit for $\lambda_B m\gg1$ to the
averaged field limit $\lambda_B m \ll 1$ is not monotonic. For instance, for
$0.25 \lesssim \lambda_B m \lesssim 0.65$, we observe the occurrence of
velocity shift maxima at the field minima and vice versa, see
Fig.~\ref{fig:map}. This can be interpreted as another manifestation of the
nonlocal nature of fluctuation-induced properties: e.g., the velocity shift in
a field minimum can be dominated by the contributions from nearby maxima if
the latter are within the scale of the fluctuation size $1/m$.  

\begin{figure}[t]
\centerline{\includegraphics[scale = 0.22]{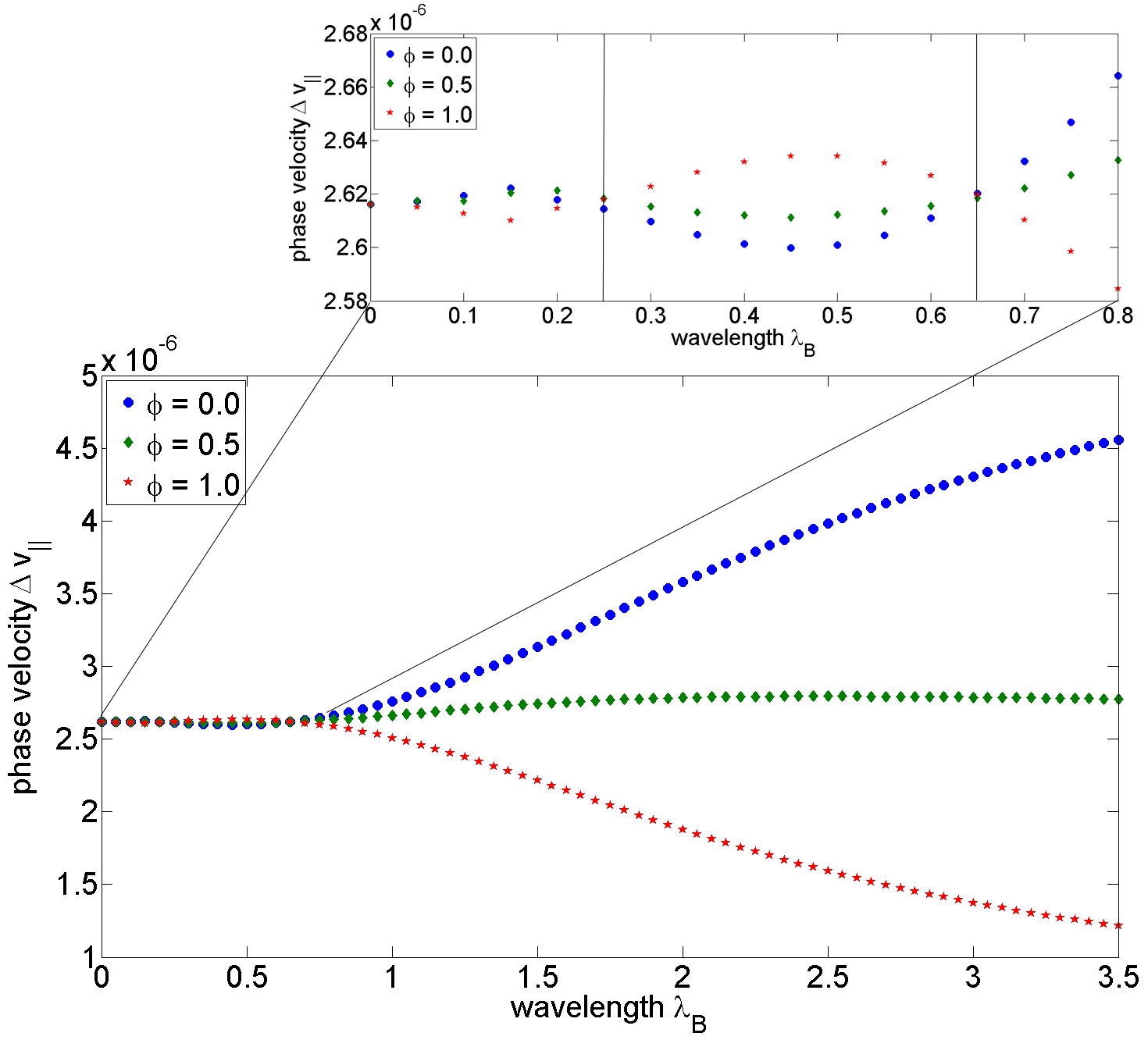} }
\caption{Phase velocity shift $\Delta v_\|$ for different lateral positions in
  the phase $\phi=2 \pi x_{3}/\lambda_B$ of the inhomogeneous-field variation
  as a function of the variation length $\lambda_B$.  The curves correspond to
  horizontal cuts of the contour plot \ref{fig:map} at
  $\phi=2x_{3}/\lambda_B=0,1/2\pi,\pi$. The standard ordering from large to
  small background field for large $\lambda_B$ from top to bottom (blue dots
  at field maximum to red stars at minimum field) can be inverted depending on
  the value of $\lambda_B$ (see inlay). The straight lines in the inlay at
  $\lambda_{B} \simeq 0.25, 0.65$ mark the inversion points.  }
\label{fig:Lines}
\end{figure}

This nonmonotonic behavior is quantitatively highlighted in
Fig.~\ref{fig:Lines}, where the phase velocity shift $\Delta v_\|$ is shown
for different positions in the phase of the variation $\phi=2 \pi
x_{3}/\lambda_B$ as a function of the variation length $\lambda_B$.
These curves correspond to horizontal cuts of the contour plot \ref{fig:map}
at the phase values $\phi= 0, 1/2\pi, \pi$.  For large $\lambda_B$, the
velocity shift approaches its constant-field limit \eqref{eq:vshift}
respecting the ordering from large to small background field. By contrast,
this ordering is reversed in the interval $0.25 \lesssim \lambda_B m \lesssim
0.65$.  For $\lambda_B\lesssim 0.25$, our data is compatible with the standard
ordering of the velocity shift in phase with the external field. In the
worldline picture, this corresponds to the next minima or maxima entering the
fluctuation average over the spatial extent of the fluctuation. A similar
phenomenon had already been observed for the case of electron-positron pair
production in inhomogeneous electric fields\cite{Gies:2005bz}. However, the
nonmonotonic signal becomes very small for small $\lambda_B$. A size of the
numerical error can be estimated from the curve for $\phi=1/2\pi$ (green
diamonds) in Fig.~\ref{fig:Lines}: owing to the symmetry of the problem, this
line should be completely flat for all $\lambda_B$. The slight deviations from
flatness thus correspond to the error of the numerical
algorithm\cite{Gies:2011he}. 

The local velocity shifts correspond to the local refractive-index shift of
the magnetized quantum vacuum. Apart from birefringence, the optical
properties of the quantum vacuum also include a self-focussing property\cite{Kharzeev:2006wg}, 
which is related to the fact that the refractive index
of the vacuum increases with increasing field strength. This statement clearly
holds for the constant-field approximation and implies that propagating
photons are bent towards local maxima of the field strength, thereby further
enhancing the field strength. Our results now show a new nonmonotonic behavior
of the refractive properties of the quantum vacuum, indicating that the
self-focussing property can be turned into a de-focussing property for rapidly
varying fields. The critical scale $\lambda_{\text{cr}}$ of field variations
where self-focussing can be converted into defocussing is given by the first
inversion point in Fig.~\ref{fig:Lines}, $\lambda_{\text{cr}}m \simeq
0.65$. Our observations thus indicate the existence of a new inherent property
of the quantum vacuum which is induced by the nonlinear as well as the
nonlocal properties of quantum fluctuations.

\section{Conclusions and Outlook}

We have argued that the photon polarization tensor deserves major attention,
as it is the central quantity for investigating and understanding
vacuum polarization effects in intense fields.  In particular recent advances
in the field of laser physics and growing interest in the search for
beyond-the-standard-model particles demand for new insights. Therefore, we
have devised two different strategies: First we focussed onto the situation of a
homogeneous field, and aimed at analytical insights into the photon
polarization tensor in the non-perturbative regime, while keeping its full
momentum dependence. These insights are particularly valuable for
  physical phenomena that need to be formulated in coordinate space, as a
  Fourier transform of course requires also information off the light cone.
Second, we employed worldline numerical methods to compute
the polarization tensor in general inhomogeneous fields. 
Both approaches provided
us with new results that can be used to tackle a variety of
problems. Whereas the polarization tensor in momentum space provides
  information analogous to the realm of geometric optics, our studies pave the
  way to also explore diffractive phenomena. Concrete applications range
from precise predictions of experimental observables in inhomogeneous fields,
to the restriction of the parameter space for beyond-the-standard-model
particles.

\section*{Acknowledgments}

B.D. and H.G. acknowledge support by the DFG under grants SFB-TR18 and
GI~328/4-1 (Heisenberg program) as well as GRK 1523.  L.R. is grateful to the
Carl-Zeiss Stiftung for financial support (PhD fellowship). We thank
J.~Jaeckel for interesting discussions and helpful correspondence.

\section{References}


\end{document}